\documentclass[twocolumn,trackchanges]{aastex631}

\begin{document}

\title{Long-term study of the first Galactic ultraluminous X-ray source Swift~J0243.6+6124 using NICER}
\author[0000-0002-9680-7233]{Birendra Chhotaray}
\affiliation{Astronomy and Astrophysics Division, Physical Research Laboratory, Navrangpura, Ahmedabad - 380009, Gujarat, India}
\affiliation{Indian Institute of Technology Gandhinagar, Palaj, Gandhinagar - 382055, Gujarat, India}

\author[0000-0002-6789-2723]{Gaurava K. Jaisawal}
\affiliation{DTU Space, Technical University of Denmark, Elektrovej 327-328, DK-2800 Lyngby, Denmark}

\author[0000-0003-3840-0571]{Prantik Nandi}
\affiliation{Astronomy and Astrophysics Division, Physical Research Laboratory, Navrangpura, Ahmedabad - 380009, Gujarat, India}

\author[0000-0003-2865-4666]{Sachindra Naik}
\affiliation{Astronomy and Astrophysics Division, Physical Research Laboratory, Navrangpura, Ahmedabad - 380009, Gujarat, India}

\author[0000-0003-0071-8947]{Neeraj kumari}
\affiliation{Astronomy and Astrophysics Division, Physical Research Laboratory, Navrangpura, Ahmedabad - 380009, Gujarat, India}

\author[0000-0002-0940-6563]{Mason Ng}
\affiliation{MIT Kavli Institute for Astrophysics and Space Research, Massachusetts Institute of Technology, 70 Vassar Street, Cambridge, MA 02139, USA}

\author[0000-0001-7115-2819]{Keith C. Gendreau} 
\affiliation{Astrophysics Science Division, 
  NASA's Goddard Space Flight Center, Greenbelt, MD 20771, USA}

\begin{abstract}
    We present the results obtained from detailed X-ray timing and spectral studies of X-ray pulsar Swift~J0243.6+6124 during its giant and normal X-ray outbursts between 2017 and 2023 observed by the Neutron star Interior Composition Explorer (NICER). We focused on the timing analysis of the normal outbursts. A distinct break is found in the power density spectra of the source. The corresponding break frequency and slope of power-laws around the break vary with luminosity, indicating the change in accretion dynamics with mass accretion rate. Interestingly, we detected quasi-periodic oscillations within a specific luminosity range, providing further insights into the underlying physical processes. We also studied the neutron star spin period evolution and a luminosity variation in pulse profile during the recent 2023 outburst. The spectral analysis was conducted comprehensively for the giant and all other normal outbursts. We identified a double transition at luminosities of $\approx$~7.5$\times$10$^{37}$ and 2.1$\times$10$^{38}$~erg~s$^{-1}$ in the evolution of continuum parameters like photon index and cutoff energy with luminosity. This indicates three distinct accretion modes experienced by the source mainly during the giant X-ray outburst. A soft blackbody component with a temperature of 0.08-0.7~keV is also detected in spectra. The observed temperature undergoes a discontinuous transition when the pulsar evolves from a sub- to super-Eddington state. Notably, in addition to an evolving 6-7~keV iron line complex, a 1~keV emission line was observed during the super-Eddington state of the source, implying the X-ray reflection from the accretion disc or outflow material.
\end{abstract}

\section{Introduction} \label{sec:intro}

Ultraluminous X-ray sources (ULXs) are non-nuclear point-like X-ray sources observed in extra-galactic regions that exhibit luminosities exceeding 10$^{39}$ erg s$^{-1}$ (for a detailed review see \citealt{2017ARA&A..55..303K}). Initially, these X-ray sources were speculated to be intermediate-mass black holes (IMBH) \citep{1999ApJ...519...89C}. Later, it was observed that the ULXs are stellar-mass black holes or neutron stars that are present in close binaries in which the donor fills the Roche lobe and accretion occurs in supercritical/super-Eddington mode \citep{2001ApJ...552L.109K}. In 2014, the first Ultraluminous X-ray pulsar (ULXP) was discovered, with coherent pulsations detected from the ULX \texttt{M82~X-2} by \citet{2014Natur.514..202B}. This finding provided a unique opportunity to study super-Eddington accretion onto the magnetized neutron stars. Since then, several ULXPs have been detected, contributing to our understanding of these systems (\citealt{2019MNRAS.484..687M} \& references therein). 

The accretion process and radiation conversion in ULXPs and typical accretion-powered X-ray pulsars share similarities. Accretion-powered X-ray pulsars possess a strong magnetic field that directs the infalling matter toward the polar regions of the neutron star. Due to the intense gravitational field near the neutron star, the accreting material gains substantial kinetic energy. Eventually, depending on the mass accretion rate, the material decelerates through the gas or radiative shocks as it reaches the neutron star's surface. The combination of accumulated material and generated radiation in that region forms an accretion column, which serves as the dominant source of radiation in accreting pulsars (\citealt{2007ApJ...654..435B, 2012A&A...544A.123B,2013A&A...551A...1R}). The primary emission in accreting pulsars consists of seed photons, in soft X-rays,  originating from the hot spots above the magnetic poles. These photons undergo Compton upscattering with the infalling plasma giving rise to broadband X-ray emission \citep{2007ApJ...654..435B}. The broadband continuum spectra are typically described by a blackbody component and a cutoff power-law component \citep{2007ApJ...654..435B,2013A&A...551A...1R}. These systems also exhibit emission features mostly from iron (Fe) and absorption features in the 10-100 keV range caused by the cyclotron resonant scattering \citep{2015A&ARv..23....2W, 2016MNRAS.461L..97J, 2019A&A...622A..61S,2023MNRAS.518.5089C}. The latter provides a direct means to estimate the magnetic field strength of the pulsar.

In addition to their spectral characteristics, accreting pulsars display various temporal properties. They exhibit coherent pulsations corresponding to their spin periods and often undergo spin-up phases, where the spin-up rate increases with increasing luminosity \citep{1973ApJ...184..271L}. The pulse profiles of accreting pulsars also show variation with the change in mass accretion rate onto the neutron star \citep{1989PASJ...41....1N, 2018ApJ...863....9W, 2023MNRAS.521.3951J}. The power density spectrum (PDS) of pulsars exhibits intriguing characteristics. It typically exhibits narrow features, indicating the presence of pulsations, as well as broad features that are attributed to quasi-periodic oscillations (QPOs) in the source \citep{1987ApJ...316..411V,2008ApJ...685.1109R,2018ApJ...863....9W}.

A remarkable transient X-ray outburst observed between 2017 and 2018 led to the discovery of Swift~J0243.6+6124. Historically, the Swift/BAT (15-50 keV) telescope detected this source at a flux level of approximately 80 mCrab on October 3, 2017, marking the onset of its 2017-2018 giant X-ray outburst \citep{2017ATel10809....1K}. Swift~J0243.6+6124 is considered to be the first Galactic ULX due to its intense X-ray luminosity reaching up to an order of 10$^{39}$~erg~s$^{-1}$ \citep{2018ApJ...863....9W,2019ApJ...885...18J,2020MNRAS.491.1857D}. Timing investigations revealed that Swift~J0243.6+6124 hosts a neutron star with a pulsation period of 9.8 seconds \citep{2017ATel10809....1K,2017ATel10812....1J,2018MNRAS.474.4432J}. Optical spectroscopic observations identified the companion (donor) star as an O9.5Ve type star~\citep{2017ATel10822....1K, 2020A&A...640A..35R}. The system is known to have a relatively short orbital period ($P_{orb}$) of around 28 days and a mildly eccentric orbit with an eccentricity ($e$) of approximately 0.1 \citep{2018A&A...613A..19D,2018ApJ...863....9W}.

A rapid spin-up rate of the pulsar was observed during the giant outburst \citep{2018A&A...613A..19D}. The pulse profile and the pulsed fraction ($PF$)  exhibited complex variation with luminosity and showed a significant change around the critical luminosity of $\sim$ 10$^{38}$ erg s$^{-1}$  \citep{2018ApJ...863....9W,2018MNRAS.479L.134T}. A QPO-like feature at 50-70 mHz is reported only in a particular luminosity range in the power density spectra (PDS) \citep{2018ApJ...863....9W}. Further timing analysis by \citet{2020MNRAS.491.1857D} revealed major changes in pulse profiles and power spectrum at specific luminosity levels. 

The broadband continuum of the source during the outburst is well described by an absorbed cutoff power law and black body components \citep{2018MNRAS.474.4432J}. A two-component transition in the values of the spectral parameters is observed during the giant outburst \citep{2020ApJ...902...18K}. Recent studies provided significant advancements in understanding the magnetic field of this pulsar. \citet{2019ApJ...885...18J} conducted a detailed investigation of the iron line width evolution with luminosity, suggesting a possible origin from the disc would require a magnetic field range of 10$^{11}$ to 10$^{12}$ Gauss. Additionally, no evidence of a cyclotron absorption line below 100 keV was found \citep{2018MNRAS.474.4432J,2021MNRAS.500..565B}. \citet{2022ApJ...933L...3K} reported the discovery of the CRSF using data from Insight-HXMT in the 120-140 keV range, estimated the magnetic field of the pulsar to be $\sim$1.6$\times$10$^{13}$ Gauss.

Swift J0243.6+6124 showed giant and subsequent normal outbursts within MJD 58029 (2017-10-03) and MJD 58533 (2019-02-19). The source also recently went into an outburst phase between MJD 60097 (2023-06-02) and 60190 (2023-09-03) (Figure~1). In this paper, we used Neutron star Interior Composition Explorer (NICER) observations for the long-term study of Swift J0243.6+6124 in the soft X-ray band.  The detailed spectral characteristics of the source in the soft X-ray band are still a matter of investigation. Thus,  a thorough spectral study of the source during the pulsar's giant and subsequent normal X-ray outbursts, including the new outburst in 2023 is performed in this paper. Moreover, our timing analysis is primarily focused on the multiple normal outbursts, including the recent outburst in 2023, during the post-giant outburst period between MJD 58303-60190. The long-term monitoring capability of NICER allows us to understand the source's spectral and timing characteristics over many magnitudes of luminosity. This paper is organized as follows: Section~\ref{sec:2} provides an overview of the observations and data reduction procedures applied to the NICER data. Section~\ref{sec:3} presents the results of the timing analysis. Section~\ref{sec:4} focuses on results obtained from the spectral analysis.The  discussions and conclusions are presented in Sections~\ref{sec:5} and ~\ref{sec:6}, respectively.

\section{ Observations \& Data Reduction}
\label{sec:2}

NICER, launched in June 2017 and installed on the International Space Station, is equipped with the X-ray Timing Instrument (XTI; \citet{2016SPIE.9905E..1HG}) designed to operate in the 0.2-12 keV energy range. The XTI comprises 56 X-ray concentrator optics, each paired with a silicon drift detector, providing non-imaging observations \citep{2016SPIE.9905E..1IP}. These concentrator optics are comprised of 24 nested grazing-incidence gold-coated aluminum foil mirrors of parabolic shape. The XTI offers a high time resolution of approximately 100 ns (rms) and a spectral resolution of about 85 eV at 1 keV. Its field of view covers approximately 30 arcmin$^{2}$ in the sky. The effective area of NICER is approximately 1900 cm$^{2}$ at 1.5 keV, utilizing 52 active detectors. The XTI is divided into seven groups of eight focal plane modules (FPMs), with each group managed by a Measurement/Power Unit (MPU) slice. This arrangement enables independent operation and control of the FPMs within each group.

\begin{figure*}
    
    \centering
    \includegraphics[trim={0 0cm 0 0},clip,scale=0.8]{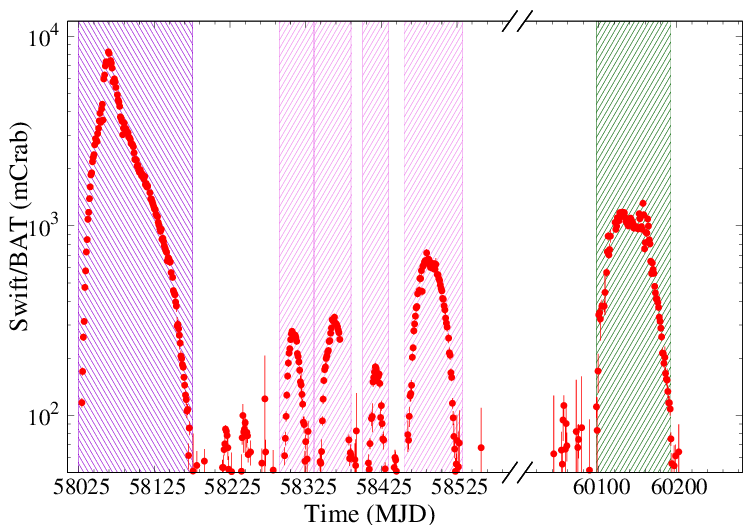}
    \caption{Swift/BAT monitoring light curve (red solid points) of the pulsar Swift~J0243.6+6124 in the 15-50 keV range, between MJD 58010 (2017-09-17) and 60220 (2023-10-03). The light curve is binned by 1 day time frame.  The outbursts that are studied using NICER observations are represented with shaded regions. The dark-violet, violet, and green color mark giant, subsequent normal outbursts, and the 2023 normal outburst phase of the source, respectively.} 
    \label{fig:Swiftlcurve}
\end{figure*}

\begin{figure}
   
    \centering
    \includegraphics[trim={0 0cm 0 0},clip,scale=0.33]{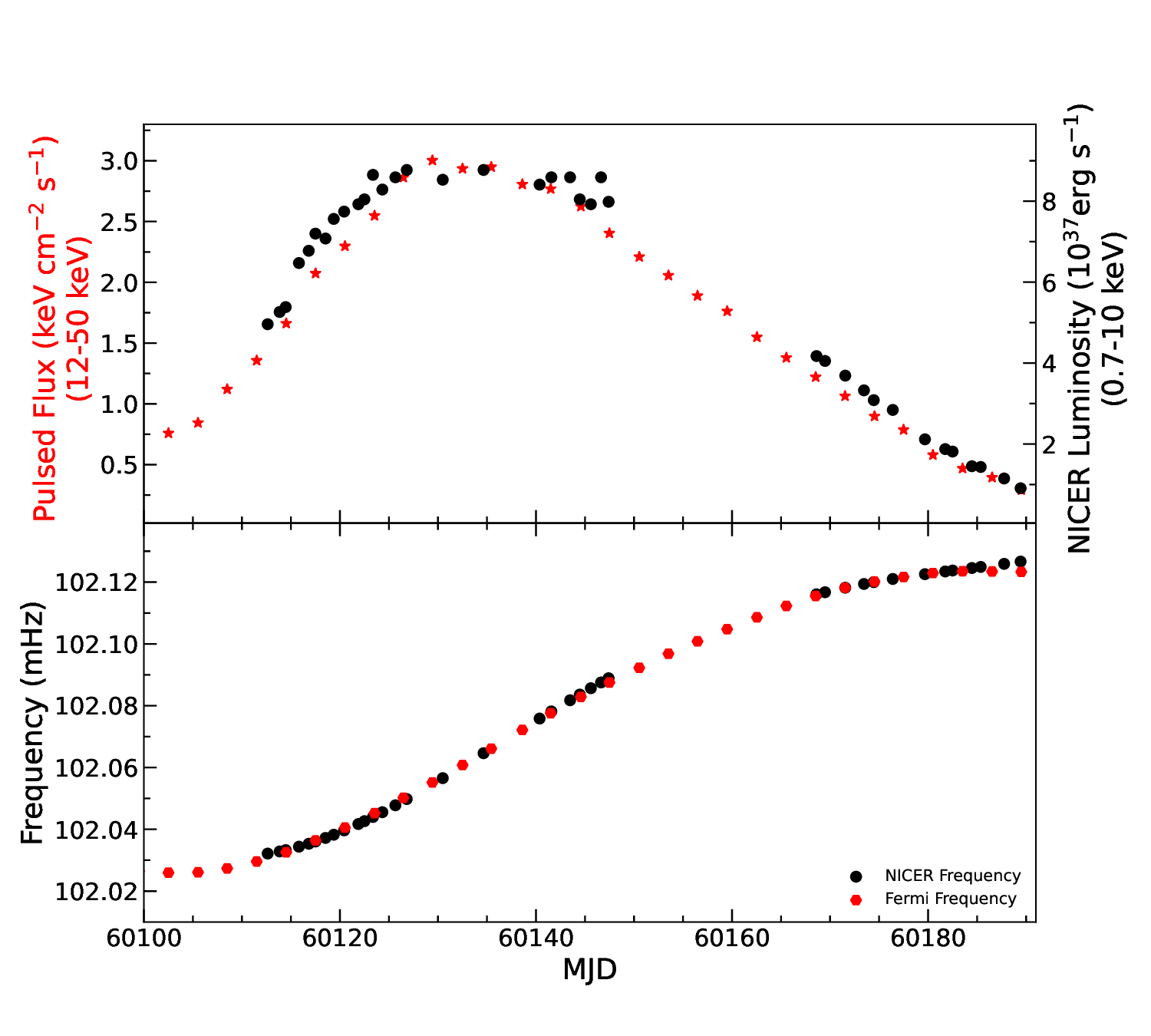}
    \caption{Upper panel: Fermi/GBM flux in the 12-50 keV band (red) and NICER luminosity in 0.7-10 keV band (black) evolution during the 2023 outburst. Bottom panel: Spin frequency evolution of Swift J0243.6+6124 measured using Fermi/GBM (red) and NICER (black) during the 2023 outburst.} 
    \label{fig:spin_evol}
\end{figure}

We used publicly available NICER data of Swift~J0243.6+6124 observed between MJD 58029 (2017-10-03) and 58533 (2019-02-19) in our study for giant and subsequent normal X-ray outbursts. These data are stored under observation ids 1050390xxx  with a net exposure time of 408 ks. Moreover, for the recent 2023 outburst, we accumulated around 104~ks net exposure data observed from June to September 2023 under observation ids 6050390227--6050390277. We used the {\tt nicerl2} pipeline available under \textsc{HEASoft} version 6.30 to process unfiltered event data of NICER. The analysis is performed in the presence of gain and calibration database files of version \texttt{xti20221001}. Good time intervals (GTI) are selected by using \texttt{nimaketime} tool alongside the application of standard filtering criteria based on the South Atlantic Anomaly (SAA), the elevation angle of $>15^\circ$ from the Earth limb, a $>30^\circ$ offset from the bright Earth, and pointing offset of 0.015$^\circ$, on the clean data after {\tt nicerl2}. The magnetic cutoff rigidity of 4~GeV/c is also applied to filter any possible high energy charge particle backgrounds. We also considered SUNSHINE==0 filtering in GTI to extract the night side data from the 2023 outburst \citep{Ng2023ATel16121....1N}. This is included due to the known visible-light leak\footnote{\url{https://heasarc.gsfc.nasa.gov/docs/nicer/}}  in the XTI optical bench of NICER on May 22, 2023. To account for the effects of Earth and satellite motion during the observations, barycentric correction is applied using solar system ephemeris \texttt{JPL-DE430}. We extracted the light curves and spectrum from each observation in \textsc{XSELECT}. The dead time correction to the light curves is not applied as the count rate is below 20000 counts~s$^{-1}$\footnote{\url{https://heasarc.gsfc.nasa.gov/docs/nicer/data_analysis/nicer_analysis_tips.html}} during these normal outbursts. However, we considered the effect of dead time on spectra during the 2017-18 giant outburst by following  \citet{2018ApJ...863....9W}.  The corresponding spectral background of each observation is obtained using the {\tt nibackgen3C50} tool \citep{2022AJ....163..130R}. The spectral response matrix and ancillary response files are created using {\tt nicerrmf} and {\tt nicerarf} commands, respectively.

Furthermore, in conjunction with NICER, we incorporated the daily X-ray monitoring data obtained from Swift/BAT. These observations encompassed the energy range of 15-50 keV, as detailed in the study conducted by \citet{2013ApJS..209...14K}. To ensure data integrity, the rows marked as good quality data, as indicated by a DATA\_FLAG value of 0, were included in our analysis.

\section{Timing analysis and results}
\label{sec:3}
We conducted timing analysis during the post-giant outbursts including the 2023 outburst with NICER. We selected these post-giant periods (MJD 58303-60190) to examine the periodic and quasi-periodic oscillations from the neutron star. The timing properties of the pulsar during the 2017-2018 giant outburst are reported in  \citet{2018ApJ...863....9W} using NICER data. Figure~\ref{fig:Swiftlcurve} illustrates the Swift/BAT monitoring light curve (solid red circles) in the 15-50 keV band, and shaded parts above it represent probed regions using NICER data.

The NICER light curves were generated using a bin size of 0.1 seconds in the 0.5-10.0 keV energy range. The timing analysis is performed on a total of 105 Ids, each with a minimum good exposure time of 1400s.  
We searched for the pulsating signal in the light curve following the $\chi^{2}$-maximization technique \citep{1987A&A...180..275L} using the \texttt{efsearch} task of \texttt{FTOOLS} package. We also applied the orbital correction to data using the orbital parameters provided by Fermi/GBM team\footnote{\url{https://gammaray.nsstc.nasa.gov/gbm/science/pulsars/lightcurves/swiftj0243.html}}. The orbital correction was done to obtain the intrinsic spin period of the neutron star that gets affected by the binary orbital modulation. In this paper, we present the spin frequency evolution of the source during the recent 2023 outburst (Figure \ref{fig:spin_evol}). The spin frequency evolution during previous outbursts such as giant and subsequent normal outbursts between MJD  58029 and 58533 is reported in \citet{2018ApJ...863....9W} and \citet{2023MNRAS.522.6115S}, respectively.

The bottom panel of Figure \ref{fig:spin_evol} shows the pulse frequency of the pulsar obtained from NICER and publicly available Fermi/GBM data from the 2023 outburst. We have presented the evolution of 12-50 keV pulsed flux from Fermi/GBM as well as the source luminosity in 0.7-10 keV observed by NICER in the top panel of the figure. The spin frequency during the recent outburst changed from $\approx$ 102.03 to 102.12 mHz in both NICER and Fermi/GBM data. 

\begin{figure}
\hspace{-0.8cm}
 \includegraphics[scale=0.55]{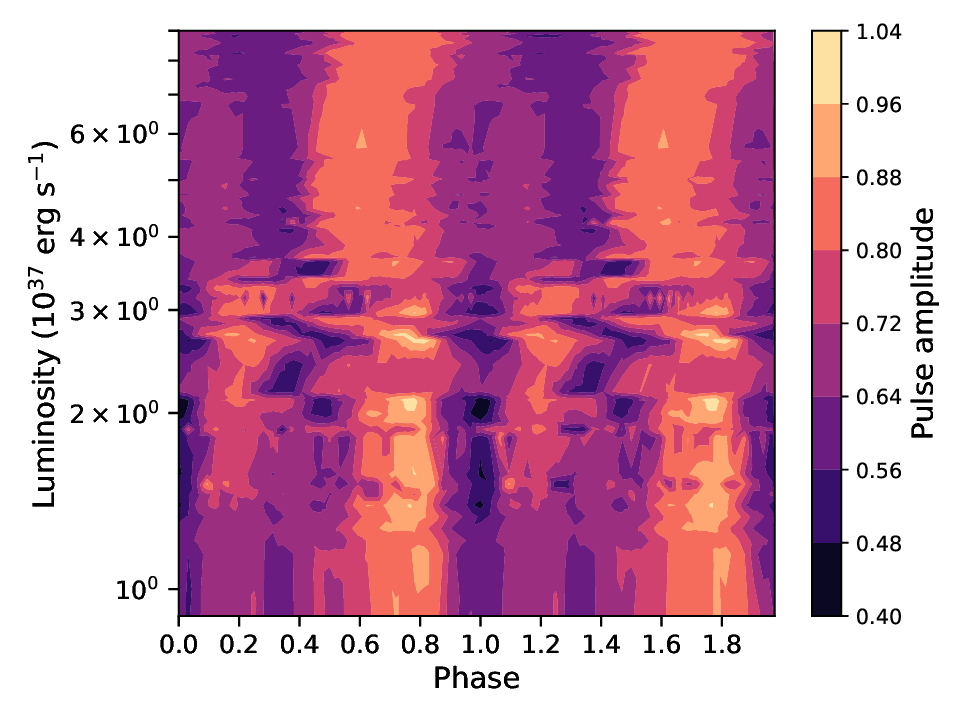}
	
 \caption{The color-coded map of the evolution of the pulse profile of the pulsar with luminosity during the normal outbursts between MJD 58303 and 60190. The pulse profiles are normalized to have values between 0 \& 1. Two cycles are shown for clarity. }
    \label{fig:pp_L_variation}
\end{figure}

\begin{figure*}
 \centering
 \includegraphics[scale=0.65, angle=-90]{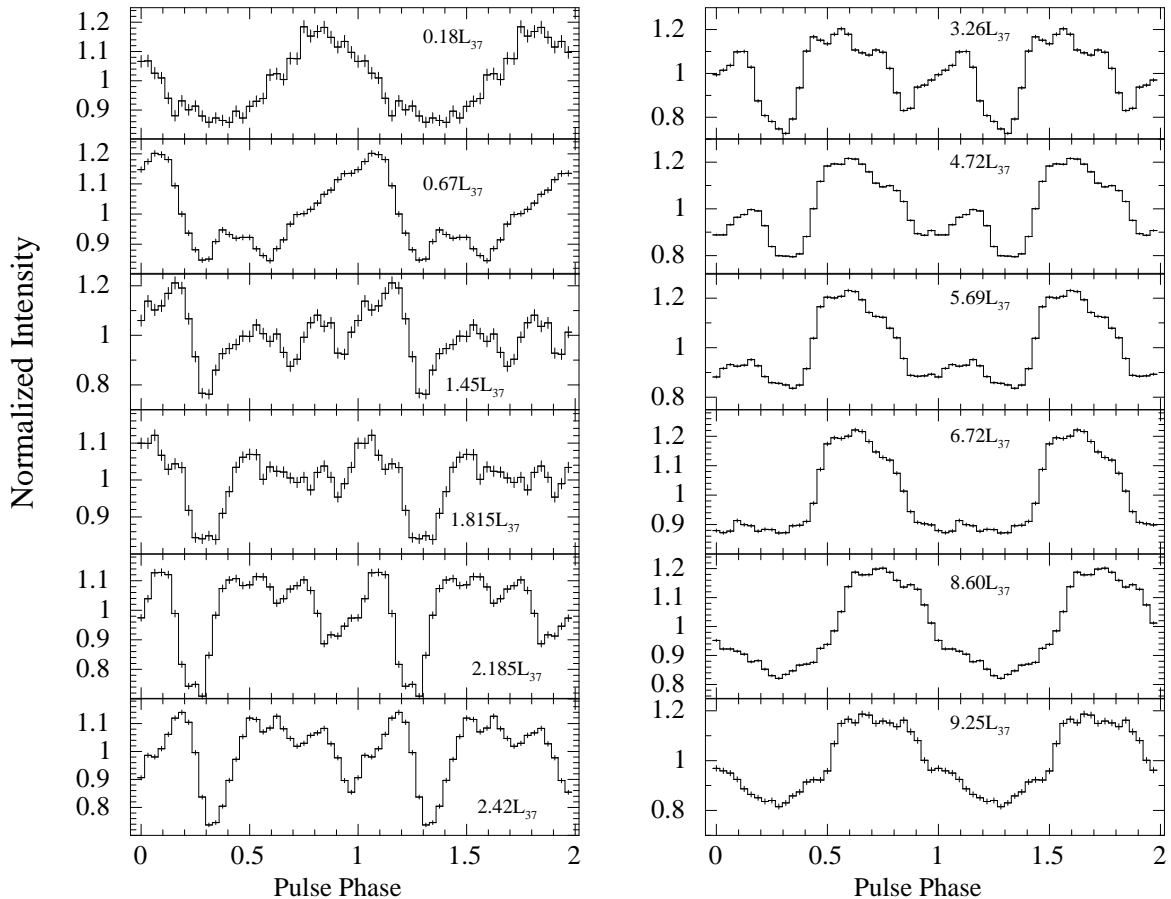}	
 \caption{
NICER pulse profiles covering a broad range of source luminosity during the normal outbursts between MJD 58303 and 60190. Two cycles are shown for clarity. Here, L$_{37}$ stands for 10$^{37}$ erg s$^{-1}$.}
    \label{fig:pp_evolution}
\end{figure*}

 We also generated the pulse profile of the pulsar during its normal outbursts including the 2023 outburst to study the geometry of pulsed beamed emission in soft X-rays.  For this, each light curve is folded at its corresponding spin period using the \texttt{efold} task of \texttt{FTOOLS} package. The pulse profile variation with luminosity from previous giant and subsequent normal outbursts observed between MJD 58029 and 58533 studied by \citet{2018ApJ...863....9W}, \citet{2023MNRAS.522.6115S}, and \citet{2023ApJ...950...42L}.  We noticed no significant difference in pulse profile evolution with luminosity during the normal outbursts between MJD 58303-58533 and 2023 normal outbursts. Hence, in Figure ~\ref{fig:pp_L_variation}, we show pulse profile evolution with luminosity during the normal outbursts between MJD 58303-58533 and 2023 normal outbursts between MJD 58303-60190 for completeness.  
 The pulse profiles appear complicated with multiple dips/notches at certain pulse phases at luminosities below 6$\times$10$^{37}$~erg s$^{-1}$. To show the detailed variation of dips/notches and profile evolution with luminosity, we represented the pulse profiles from 12 NICER observations in Figure~\ref{fig:pp_evolution}. Below the luminosity of $\sim$0.5$\times$10$^{37}$ erg s$^{-1}$, the pulse profiles are  single peak dominated in nature. Various dip-like structures arise in the pulse phase range of 0.5-1.0 for luminosity between $\sim$(0.5-2.5)$\times$10$^{37}$~erg s$^{-1}$. One of the dips became deeper, and the profiles evolved into a double-peaked structure at a luminosity $\sim$2.5$\times$10$^{37}$ erg s$^{-1}$. Further, a smooth single peaked profile is observed at luminosities beyond $\sim$6$\times$10$^{37}$ erg s$^{-1}$.

\begin{table}
    \centering
    \caption{Observed QPO frequency, width, and its significance with luminosity. The uncertainties are presented at the 1~$\sigma$ confidence level.}
    \begin{tabular}{|c|c|c|c|c| }
\hline
MJD & Luminosity &  $QPO_{f}$ & $QPO_{W}$ & Significance   \\
    & 10$^{37}$ erg s$^{-1}$ & (mHz) & (mHz) &  \\
\hline
58322 & 0.69 & 40$\pm$2 & 7$\pm$4 & 3.99 \\
58323 & 0.63 & 47$\pm$2 &12.0$\pm$2& 6.52 \\ 
58325 & 0.51 & 34$\pm$1 & 9$\pm$2 & 8.42 \\ 
58331 & 0.29 & 31$\pm$2 & 7.7$\pm$2 & 5.68 \\
58344 & 0.34 & 30$\pm$1 & 8.2$\pm$2 & 8.58 \\
58425 & 1.83 & 40$\pm$1.2 &6.50$\pm$3 & 6.52 \\
58430 & 0.49 & 38$\pm$2 &19$\pm$4 & 10.0 \\
58449 & 0.25 & 34$\pm$2 & 4$\pm$1.5 & 4.54 \\
58460 & 0.47 & 32$\pm$3 &9$\pm$2 & 4.51  \\
58521 & 0.58 & 40$\pm$1 & 5$\pm$2 & 5.36 \\
\hline
    \end{tabular}
    \label{tab:QPO_significance}
\end{table}

Furthermore, we computed the pulsed fraction ($PF$) of the pulse profiles from the 2023 outburst to understand the evolution of soft X-ray emission of the pulsar. Such a study has not been performed using NICER, except for the 2017-2018 giant outburst. Thus, we also studied the remaining post-giant outburst data in our study.  We calculated the $PF$ using the root mean squared method as given by \citet{2018ApJ...863....9W} (see also \citealt{2023A&A...677A.103F}). It is expressed as: \newline

\begin{equation}
   PF = \frac{(\sum_{i=1}^{N} (r_{i}-\overline{r})^{2}/N)^{1/2}}{\overline{r}}
\end{equation}
\newline

The variation of $PF$ with luminosity is presented in the panel (a) of Figure~\ref{fig:QPO_evolve_L}.  We fitted a curve to the obtained $PF$ values at different luminosities using the spline interpolation method, to understand the trend of $PF$ evolution. The red line is the best-fit spline curve, and the shaded area indicates the moving average standard deviation of data points. The $PF$ varied between $\sim$10\% to 15\% for the source luminosity in the range of $2\times10^{36}$ to $9\times10^{37}$ erg s$^{-1}$. A relatively higher value of $PF$ ($\sim$20-25\%) was detected at a luminosity of $\sim$$2\times10^{37}$~erg s$^{-1}$.

\subsection{Power Density Spectrum (PDS) Analysis}
 
The evolution of the PDS and its features during normal outbursts including the 2023 outburst is investigated for the first time in this work. We performed PDS analysis on 105
light curves in the 0.5-10 keV energy range using NICER data to follow the luminosity dependency variation of the PDS. Additionally, we searched for the sign of any QPO in the PDS to compare these features with the same observed during the giant outburst \citep{2018ApJ...863....9W}.

 We used the  \texttt{powspec} tool from the XRONOS package to generate the power density spectra.   The light curves were divided into intervals of $\sim$400 s, and then the  PDS of each interval was generated. The final PDS was obtained by averaging the PDS of segmented light curves. This method improves the detection probability of QPO-like features. We also employed the \texttt{powspec norm=-2} command to obtain white-noise subtracted averaged PDS.  Following this, the power is expressed in units of (RMS/mean)$^{2}$/Hz. The resulting PDS for the observation ID 1050390170 (MJD 58322) is illustrated in Figure~\ref{fig:QPO_detection}. 

\begin{figure*}
    \centering
    \includegraphics[trim={0.0cm 0.0cm 0 0.5cm},clip,scale=1.1]{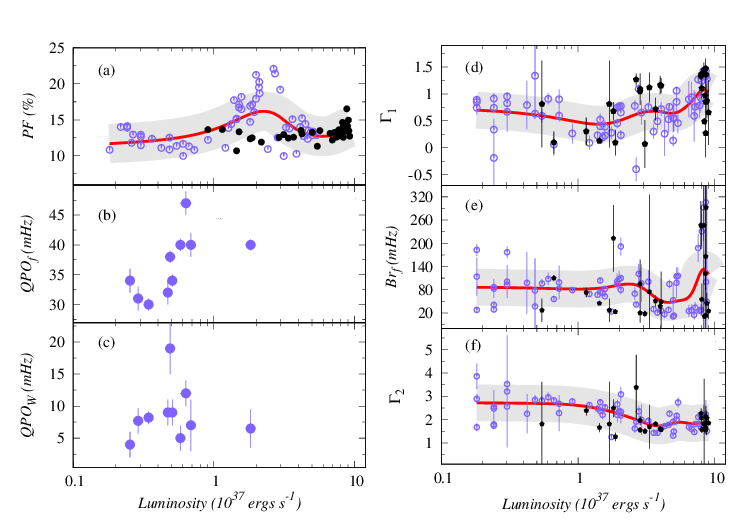}
    \caption{ 
    The luminosity evolution of different parameters from PDS analysis during normal outbursts.
    Panels (a-f) display the values of pulsed fraction ($PF$), break frequency ($Br_{f}$), QPO frequency ($QPO_{f}$), QPO width ($QPO_{W}$), slope of power-law before $Br_{f}$ ($\Gamma_1$), and slope of power-law after $Br_{f}$($\Gamma_2$) as functions of luminosity. The red line represents the best-fitting curve for these values using the spline interpolation method. The shaded area indicates the moving average standard deviation of data points. The purple data points represent values obtained from normal outbursts, except for the 2023 outburst that is shown in black. 
    }
    \label{fig:QPO_evolve_L}
\end{figure*}

The PDS exhibited narrow peaks at multiple of a main frequency around 0.101 Hz. These peaks correspond to the pulsar's spin frequency and its harmonics, which are ignored during the fitting of PDS \citep{2010MNRAS.407..285J}. Initially, we attempted to fit the PDS continuum by a simple power law model (\texttt{powerlaw}) using \texttt{XSPEC}. However, this model proved inadequate in fitting the overall PDS across a wide frequency range, spanning from approximately 0.001 Hz to 5.0 Hz. Subsequently, we replaced the power law with the broken power-law model (\texttt{bknpower}), which resulted in an acceptable chi-square ($\chi^2$) value. We studied the evolution of the slope of power-law before break ($\Gamma_{1}$), break frequency ($Br_f$),  and slope of power-law after break ($\Gamma_{2}$) in the PDS w.r.t luminosity, which is presented in the panel~(d), (e), and (f) of Figure~\ref{fig:QPO_evolve_L}, respectively. The red line represents the best-fitted curve using the spline interpolation method to examine the evolution of the parameter.  From the figure, the break frequency $Br_f$ varied  between 50 and 80~mHz below 7.5$\times$10$^{37}$ erg~s$^{-1}$. Beyond this luminosity, the  $Br_f$ increases to 140~mHz, surpassing the pulsar frequency (101 mHz). $\Gamma_{1}$ and $\Gamma_{2}$ values varied between 0.5-1 and 2-3, respectively, within the studied luminosity range.

\begin{figure}
    \includegraphics[trim={0 0.0cm 0 0},clip,scale=0.7, angle =0]{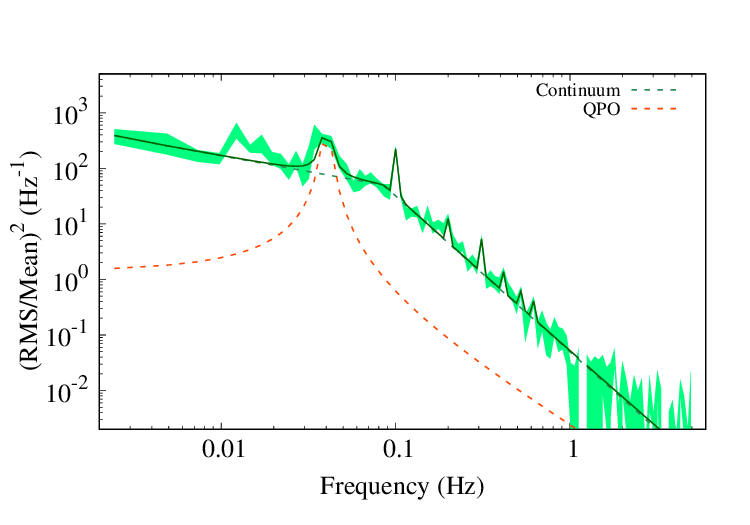}
    \caption{  
A representative power density spectrum was obtained from a NICER observation ID 1050390170. The continuum is fitted with a broken power law, and the QPO feature is described using a Lorentzian function.  The observed narrow peaks correspond to the pulsar rotational frequency and its harmonics.}
    \label{fig:QPO_detection}
\end{figure}

After fitting the PDS continuum with a broken power-law model, we examined the residuals for the presence of potential QPOs.  In addition to the neutron star's spin frequency and its harmonic components, we observed a broad hump-like residual below the pulsar spin frequency in a few cases (around 16 IDs). To explore this further, we employed a Lorentzian function (\texttt{lorentz}) as the shape of this hump appeared asymmetric. The Lorentzian function has been widely employed in several QPO studies (see e.g., \citealt{2002ApJ...572..392B}). Figure~\ref{fig:QPO_detection} displays the best-fitted model for the observation ID~1050390170 (MJD~58322) in the PDS.

Furthermore, the significance of the QPOs is determined using the method described in \citet{2010MNRAS.401.1290B} based on the Lorentzian fitting. We found that QPOs from  10 out of 16 IDs  between MJD 58322--58521 exhibited a significance of more than $3\sigma$ using the above method. The frequency (QPO$_{f}$) and full width at half maximum (QPO$_{W}$) of these detected QPOs and their respective significance are presented in Table~\ref{tab:QPO_significance}. The evolution of QPO$_{f}$ and QPO$_{W}$ with luminosity is also presented in panels (b)  and (c)  of Figure~\ref{fig:QPO_evolve_L}, respectively. We did not observe any QPO-like feature during the recent 2023 outburst of Swift J0243.6+6124.

\section{Spectral Analysis And Results}
\label{sec:4}

We conducted the spectral analysis using NICER data across multiple outbursts observed between 2017 and 2023. This allowed us to probe changes in the spectral parameters over a broad range of luminosities. The spectral analysis was performed in the energy range of 0.7-10.0 keV using the \texttt{XSPEC} (v-12.11.0, \citealt{1996ASPC..101...17A}) package. The above energy range is selected to avoid the spectral uncertainties below 0.4 keV and above 10.0 keV, and a strong edge-like feature appears near 0.5~keV, especially in the brighter observations. We found the 0.5 keV feature depends on the choice of the photo-electric absorption model as well as assumed abundances up to some extent. The 0.5~keV feature may also have a calibration origin from Oxygen edge\footnote{\url{https://heasarc.gsfc.nasa.gov/docs/nicer/analysis_threads/arf-rmf/}}. A systematic uncertainty of 1.5\% was also applied, as recommended by the instrument team. For quantifying line-of-sight X-ray absorption, \texttt{wilm} abundance table \citep{2000ApJ...542..914W}  is used with \texttt{Vern}\footnote{\url{https://heasarc.gsfc.nasa.gov/xanadu/xspec/manual/node120.html}} photo-ionization cross section. The spectra were binned with a minimum of 30 counts per energy bin to allow for the application of chi-square statistics in the analysis.

\begin{figure}
    \hspace{-0.9cm}
    \includegraphics[trim={0 0.5cm 0 0}, scale=1.3]{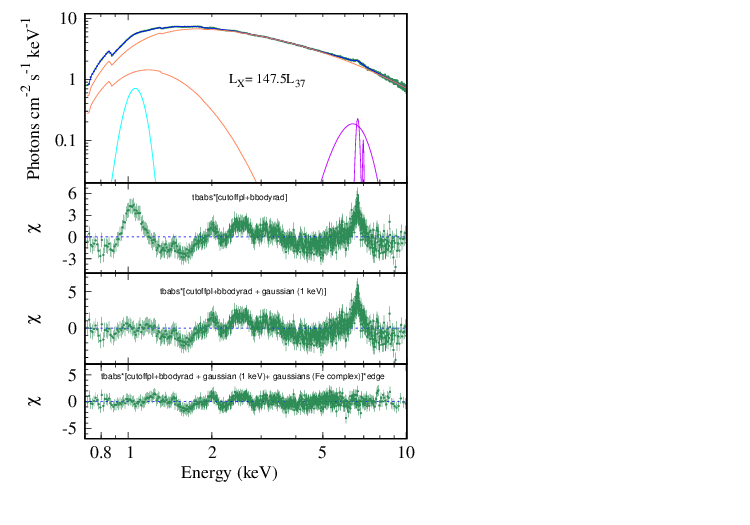}
    \caption{  The 0.7-10 keV energy spectrum of Swift~J0243.6+6124 obtained from the NICER observation on MJD 58065 (ID~1050390115) near the peak of the X-ray outburst. The second, third, and fourth panels from the top show the evolution of residual after fitting the continuum, and subsequent addition of 1 keV,  6-7 keV Fe-line complex, and edge feature, respectively.} 
   \label{fig:114id}
\end{figure}

To understand the evolution of the pulsar emission over multiple outbursts, we considered a uniform continuum model to assess the homogeneous changes in parameters with the source luminosity. 
We used absorbed cutoff power law  (\texttt{tbabs$\times$cutoffpl}) to describe the continuum emission following the previous studies \citep{2019ApJ...885...18J, 2020ApJ...902...18K}. Following the continuum model, the prominent positive residuals in the energy ranges of 6-7 keV and 0.9-1.1 keV were detected, mainly during the 2017-2018 giant outburst. Moreover, we detected $\approx$ 7.1~keV iron edge feature only during the high luminosity phases. To account for the positive residuals in the 6-7 keV range, we employed one to three Gaussian components depending on the source luminosity, as recommended in \citet{2019ApJ...885...18J}. The residuals in the 0.9-1.1 keV band were modeled using a single Gaussian component. Figure \ref{fig:114id} shows representative spectral modeling of pulsar energy emission and spectral residuals after the continuum and line emission components in different panels, obtained from a NICER observation near the peak of the giant outburst in 2017 around MJD~58065.

\begin{figure*}
    \centering
    \includegraphics[trim={0 1cm 0 0}, scale=1.2]{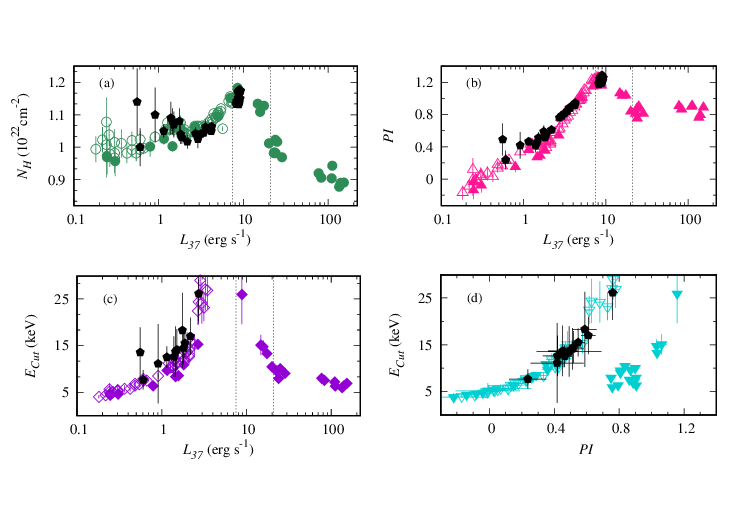}
    \caption{  The panels (a)-(c) show the evolution of the parameters such as column density ($N_{H}$), Photon Index ($PI$), and Cutoff energy ($E_{cut}$)  with luminosity obtained from multiple outbursts of Swift~J0243.6+6124. Panel (d) shows the relationship between the photon Index ($PI$) and cutoff energy ($E_{cut}$). The filled and open markers represent data points obtained from giant and subsequent normal outbursts, respectively.  The black points represent the values obtained from the 2023 outburst. 
    }
    \label{fig:spec_evolve_L}
\end{figure*}

\begin{figure*}
    \centering
    \includegraphics[trim={0 0cm 0 0}, scale=1.2]{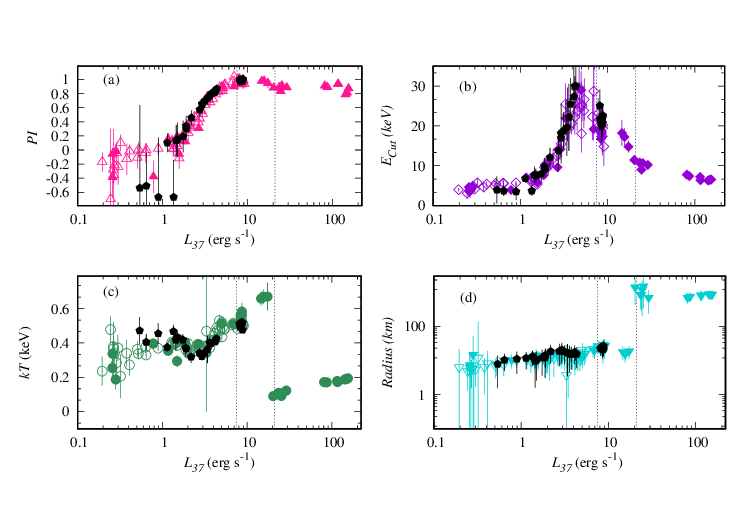}
    \caption{ The panels (a)-(d) show the luminosity dependencies of spectral parameters such as photon index ($PI$), cutoff energy ($E_{cut}$), blackbody temperature ($kT$), and blackbody radius at a fixed column density over multiple outbursts of Swift~J0243.6+6124. The symbols have the same meaning as  Figure~\ref{fig:spec_evolve_L}.
    }
    \label{fig:spec_evolve_L2}
\end{figure*}

The continuum parameters obtained from the best-fitted models on all NICER observations from multiple outbursts are presented in Figure~\ref{fig:spec_evolve_L}.
The variation of spectral parameters such as hydrogen column density ($N_{H}$), photon index ($PI$), and cutoff energy ($E_{cut}$), with unabsorbed luminosity, is shown in panels (a), (b), and (c) of Figure~\ref{fig:spec_evolve_L}, respectively. The uncertainties in the parameter values were calculated within the 90\% confidence range. Subsequently, the luminosity was estimated from the unabsorbed flux in the 0.7-10.0 keV energy range, assuming a distance of 7 kpc \citep{2018ApJ...863....9W}.

Furthermore, we noticed the $N_{H}$ varies in a narrow range between (0.9-1.2)~$\times$~10$^{22}$ cm$^{-2}$ across these observations. Therefore, to restrain any spectral degeneracies, we refitted the spectra using an absorbed cutoff power-law model at a fixed average value of $N_{H}$=1.064$\times$ 10$^{22}$ cm$^{-2}$. Following this approach, a positive residual near the lower energy side arises that could be described with a soft \texttt{bbodyrad} component. The luminosity variations of the latter model parameters obtained from an absorbed cutoff power-law model with a blackbody component are shown in Figure~\ref{fig:spec_evolve_L2}.

By examining the overall behavior of these parameters from  Figures~\ref{fig:spec_evolve_L} \& ~\ref{fig:spec_evolve_L2}, two transition points can be identified. The photon index ($PI$) exhibited a  two-component transition with luminosity that may be associated with the changes in the accretion mode (Figure~\ref{fig:spec_evolve_L}). The first transition occurred around a luminosity $L_{1}$ of  7.5$\times$10$^{37}$ erg s$^{-1}$, where $PI$ showed an increasing trend with luminosity. A decrease in the photon index was observed up to a luminosity of 2.1$\times$10$^{38}$ erg s$^{-1}$ ($L_{2}$) which can be identified as a second transition point. The photon index remains almost stable beyond $L_{2}$.  A similar trend can also be seen in Figure~\ref{fig:spec_evolve_L2}. The transitional luminosities $L_{1}$ and $L_{2}$ are marked with dotted lines in these figures.   

Moreover, we found the cutoff energy ($E_{cut}$)  increases with luminosity clearly below the first transition point $L_{1}$ (Figures~\ref{fig:spec_evolve_L} \& ~\ref{fig:spec_evolve_L2}). Due to the limited bandpass of NICER, an upper limit of 30 keV was imposed on the cutoff energy during fitting. This is in line with the maximum value observed in this pulsar based on broadband spectral analysis using HXMT data \citep{2020ApJ...902...18K}.  Between $L_{1}$ and $L_{2}$, the $E_{cut}$ steeply decreased with luminosity, whereas a gradual evolution is observed above $L_{2}$. Furthermore, we would like to highlight the unique evolution of $PI$ \& $E_{cut}$ with luminosity observed beyond the second transition point in our study. These parameters take a distinct trajectory compared to the trends reported in Figure~3 of \citet{2020ApJ...902...18K} in the super-Eddington regime. 
While \citet{2020ApJ...902...18K} identified a positive correlation between $PI$ \& $E_{cut}$ and luminosity, our findings indicate a constant $PI$ and an anti-correlation trend for $E_{cut}$ with luminosity. It is important to note that in the super-Eddington phase, the emission from an outflow or a reflection component can alter the shape of the X-ray continuum. Such an effect may not be tracked alone with NICER due to its limited energy coverage. 
Therefore, we can expect a slight variation in parameters in the super-Eddington regime. Additionally, to examine this, we conducted spectral fitting by introducing an additional blackbody component during the super-Eddington phase, as suggested by \citet{2019ApJ...873...19T}. The temperature of this blackbody component exhibited an evolution between $\approx$1  and 3 keV, potentially stemming from the contributions of the hotspot and top of the accretion column, with a radius between 10-40 km. In the presence of the second blackbody component, the photon index and cutoff energy show a positive pattern after the second transition point, similar to  \citet{2020ApJ...902...18K}. However, it is crucial to exercise caution in interpreting these findings due to the constraints imposed by the limited energy band of NICER.

In panels (c) and (d) of Figure \ref{fig:spec_evolve_L2}, the evolution of blackbody temperature and its corresponding emission radii are presented. Below $L_{2}$, the blackbody temperature ($kT$) gradually varies in a range of 0.1-0.7~keV. The size of the emission site also changes between 10 to 30 km, for the source distance of 7 kpc, below $L_{2}$. A sudden change in these parameters is observed around $L_{2}$.

\begin{figure}
    \centering
    \includegraphics[scale=1.1]{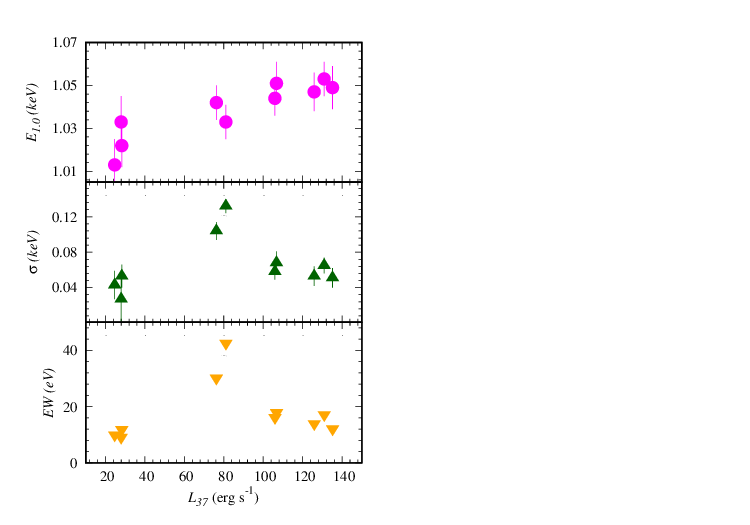}
    \caption{The figure illustrates the evolution of Gaussian model parameters of 1.0 keV  emission line. The parameters include the central line energy ($E_{1.0}$), width ($\sigma$), and equivalent width ($EW$), presented from top to bottom, respectively.}
    \label{fig:0p7_1keV_line}
\end{figure}

In addition to evolving iron emission lines in the 6-7 keV band \citep{2019ApJ...885...18J}, an emission line around 1 keV is detected in the spectra. The 1 keV feature is mainly detected during the giant outburst at luminosities above 2$\times$10$^{38}$ erg s$^{-1}$ i.e. the second transition point. The parameters obtained from the emission line analysis are presented in Figure~\ref{fig:0p7_1keV_line}. From this figure, it can be observed that the variation of central energy with luminosity falls within the error bars. However,  the line width ($\sigma$) and equivalent width ($EW$) of the 1 keV line show an increase as the luminosity increases up to $\sim$8$\times$10$^{38}$ erg s$^{-1}$, and beyond that decreases with luminosity. This indicates a strong luminosity dependency of the 1 keV emission line.

\section{Discussion}
\label{sec:5}

We have studied the first Galactic ultraluminous X-ray pulsar Swift~J0243.6+6124 using NICER observations covering a giant and multiple normal X-ray outbursts between 2017 and 2023. Such coverage allows us to probe the timing and spectral properties of this unique pulsar depending on the mass accretion rate. Our analysis over multiple outbursts revealed a wide luminosity variation in a range of (0.1-153)$\times$10$^{37}$ erg s$^{-1}$ in 0.7-10 keV energy range, assuming a distance of 7 kpc. The discussion on timing results obtained from normal outbursts is presented first in this section. We thereafter discuss the implications of observed two spectral transitions from the source. 

\subsection{Temporal characteristics of Swift~J0243.6+6124 in post-giant outburst phases}

We examined the spin frequency evolution of the pulsar during its recent 2023 outburst  (Figure~\ref{fig:spin_evol}). The spin frequency evolution from previous outbursts has been reported by  \citet{2018ApJ...863....9W} and \citet{2023MNRAS.522.6115S}. Before the 2023 outburst, the source was in a quiescent state between MJD 58535-60097, where a decrease in spin frequency was observed from 102.10 to 102.03 mHz as per the Fermi/GBM. However, the source gradually spun up to 102.12 mHz after the 2023 outburst (Figure \ref{fig:spin_evol}). This value is similar to the spin frequency of the neutron star observed after the giant and subsequent normal outbursts between 2017-2019 \citep{2018ApJ...863....9W,2023ApJ...950...42L, 2023MNRAS.522.6115S}. A significant amount of mass transfer is expected to enhance the spin frequency due to the transfer of angular momentum to the neutron star.

We further have investigated the pulse profile of the pulsar from post-giant outbursts including its recent 2023 outburst  (Figures~\ref{fig:pp_L_variation}). The pulse profiles offer insights into the geometry of the emission regime on the neutron star surface. The pulse profiles of accretion-powered pulsars tend to be simpler and smoother in hard X-rays. However, the soft X-ray pulse profiles appear complex due to the influence of circumstellar scattering and absorption \citep{1983ApJ...270..711W}. In our analysis, we observed complex pulse profiles in the 0.5-10 keV energy band , featuring various dips or notches at different pulse phases (see Figure~\ref{fig:pp_evolution}). Similar features have been observed in the pulse profiles of other Be/X-ray binary pulsars such as V0332+53, 1A~0535+262, EXO~2030+375, GX~304-1, and RX~J0209.6-7427 \citep{2006MNRAS.371...19T, 2008ApJ...672..516N,2013ApJ...764..158N, 2015RAA....15..537N, Epili2017MNRAS.472.3455E, Jaisawal2016MNRAS.457.2749J, 2020MNRAS.494.5350V}. These features are usually attributed to the absorption of photons by matter streams locked at specific pulse phases of the neutron star, demonstrating the dynamics of matter distribution in the magnetosphere.

The critical luminosity for Swift~J0243.6+6124 can be considered to be $\approx$10$^{38}$~erg~s$^{-1}$ based on the evolution of the source during its giant outburst \citep{2018ApJ...863....9W}. Assuming this limit, the pulsar was accreting in a sub-critical or close to critical luminosity regime during the normal outbursts between MJD 58303-58533 and the recent 2023 outburst where the observed luminosity was in the range of (0.2-9.0)$\times$10$^{37}$~erg~s$^{-1}$.  We presented the pulse profile evolution of the pulsar with luminosity during the normal outbursts in Figure~\ref{fig:pp_L_variation}. Initially, the pulse profiles are  single peak dominated in nature when the luminosity was below $\sim$0.5$\times$10$^{37}$ erg s$^{-1}$. Between $\sim$0.5-6$\times$10$^{37}$erg s$^{-1}$, various dips/notches arise and the profile evolves to double-peaked. A clear double-peaked profile is visible below the first transition point $L_{1}$. Moreover, the pulse profile evolved from a double-peaked to a smooth single-peaked structure at luminosities above $\sim$6$\times$10$^{37}$ erg s$^{-1}$,  which is similar to the pulse profiles shape reported by \citep{2018ApJ...863....9W} around this luminosity. The observed pulse profile evolution even below the critical regime suggests a change in the emission geometry or beam pattern of the neutron star depending on the mass accretion rate.

In addition to the pulse profile, the pulsed fraction ($PF$)  provides valuable insights into the pulsating emissions originating from a source. We observed a variation in the $PF$ corresponding to changes in luminosity (Figure~\ref{fig:QPO_evolve_L}). The pulsed fraction ($PF$)  exhibits a moderate variation from  $\sim$ 10\% to 15\% within the luminosity range of  $\sim2\times10^{36}$ to $\sim9\times10^{37}$ erg s$^{-1}$.  
Notably, during the giant outburst, \citet{2018ApJ...863....9W} found an increase in the pulsed fraction with rising luminosity above a critical luminosity of 10$^{38}$ erg s$^{-1}$ from 20\% to 55\%. Below the critical luminosity, the $PF$ varies between 15-30\% during the giant outburst. This is similar to our present findings during the post-giant outbursts. Based on the examination of $PF$ and observed outburst luminosities, the source was accreting below or close to the critical regime at the peak of these multiple normal outbursts.

\subsection{Detection of QPO and break feature in the PDS}

We observed low-frequency QPOs from Swift~J0243.6+6124 at particular epochs within a given range of luminosity during the subsequent normal outbursts after the giant outburst. The detected QPO frequencies range from 30 to 47 mHz within a luminosity range of (0.2-2.0)~$\times$~10$^{37}$~erg~s$^{-1}$ (see Table~\ref{tab:QPO_significance}).  In panels (b) and (c)  of Figure~\ref{fig:QPO_evolve_L}, we show the evolution of QPO frequency and width with the luminosity, respectively. The QPO frequency appears to be positively related to luminosity overall, however, the width remains almost flat. We detected no QPO-like signatures in the 2023 X-ray outburst with NICER.  No QPO was also found in the NuSTAR observation during the 2023 outburst \citep{2023ATel16139....1P}. During the 2017-18 giant outburst of Swift~J0243.6+6124, the QPOs with frequencies between 50-70 mHz within the luminosity range of 0.28-2.05$\times$~10$^{37}$~erg~s$^{-1}$ in 0.2-12 keV was observed with NICER by \citet{2018ApJ...863....9W}. Moreover, the power spectra obtained with Insight-HXMT in the 20-40 keV range revealed weak QPOs with luminosity-dependent frequencies ranging from $\approx$50 to 200 mHz during the giant X-ray outburst \citep{2020MNRAS.491.1857D}.   This shows the transient nature and energy dependency of the low-frequency QPOs. The detected QPO frequencies in our study are consistent with the previous QPO studies in other high mass X-ray binary (HMXB) accreting pulsars \citep{2011BASI...39..429P}. Also, the phenomenon of QPO detection primarily in the low luminosity phase has been observed in other cases, such as in KS~1947+300 \citep{2010MNRAS.407..285J}.

In the case of HMXBs, the physical origin of these QPOs is commonly explained by two models: the Keplerian Frequency Model (KFM; \citealt{1987ApJ...316..411V}) and the Magnetospheric Beat Frequency Model (MBFM; \citealt{1985Natur.316..239A}).  The QPO feature arises when the accretion process is governed by the interaction between the co-rotating magnetosphere and the inhomogeneities in the inner accretion disc, which creates variabilities in the mass accretion rate. KFM states this variable mass accretion rate occurs at the Keplerian frequency. For MBFM, this happens at a beat frequency between pulsar spin frequency and matter rotational frequency at the inner disc. In the case of Swift~J0243.6+6124, the spin frequency of the pulsar is approximately 102.1 mHz, while the detected QPO frequency ranges from 30 to 47 mHz, nearly one-third of the pulsar frequency. When the spin frequency of the pulsar exceeds the Keplerian frequency at the inner edge of the accretion disc, the co-rotating magnetosphere throws away the accreted matter outwards, which is also called centrifugal inhibition of accretion \citep{1973ApJ...184..271L}. The KFM is therefore applicable only when the QPO frequency is higher than the neutron star spin frequency, as observed in transient Be/X-ray binary pulsars like EXO 2030+375 \citep{1989ApJ...346..906A} and 1A 0535+262 \citep{1996ApJ...459..288F}.   Therefore, the MBFM could potentially explain the origin of the quasi-periodic variabilities in Swift~J0243.6+6124, which applied in previous studies like 4U 0115+634 \citep{2013MNRAS.434.2458D} and KS 1947+300 \citep{2010MNRAS.407..285J}.

Based on the detected QPO frequency, we can also calculate the magnetic field of the neutron star using the MBFM. The magnetic field of the neutron star for a dipolar structure is given by  \begin{equation}\label{eq:q}
  B =  4 \times 10^{11}  M_{1.4}^{-5/6}R_{6}^{-3}\dot{M}_{-8}^{1/2}(\frac{\Omega_{K}}{mHz})^{-7/6} ~\text{G}
  \end{equation}
Where, $M_{1.4}$=1.4$M_{\odot}$ (mass of the neutron star), $\dot{M}_{-8}$=10$^{-8}M_{\odot}$ yr$^{-1}$ (mass accretion rate), and $R_{6}$=10$^{6}$ cm (radius of neutron star).

According to MBFM, $\Omega_{K}$ = $\Omega_{QPO}$ +$\Omega_{s}$, where $\Omega_{K}$, $\Omega_{QPO}$, and $\Omega_{s}$ are the Keplerian frequency, QPO frequency, and pulsar spin frequency respectively.  Within the luminosity range of 0.2-2.0 $\times$ 10$^{37}$ erg s$^{-1}$ (assuming a distance of approximately 7 kpc), the mass accretion rate ($\dot{M}$) is estimated to be between 0.02-0.13 $\times$ 10$^{-8}$ M$_{\odot}$ yr$^{-1}$. By applying these values in Equation~\ref{eq:q} and considering the detected QPO and pulsar frequencies, the magnetic field ($B$) of the neutron star is estimated to be approximately 2.5 $\times$ 10$^{12}$ G.  The obtained magnetic field is in good agreement with measurements provided by previous studies using indirect methods~\citep{2018MNRAS.479L.134T, 2020MNRAS.491.1857D, 2022MNRAS.516.1601B}. For example, the non-detection of the propeller effect by \citet{2018MNRAS.479L.134T} constrained the upper limit of the magnetic field to 6$\times$10$^{12}$~G. Additionally, \citet{2020MNRAS.491.1857D} provided a range of magnetic field values after analyzing various features such as propeller effect, state transition luminosity from sub- to super-critical state, transitional luminosity from gas pressure dominated (GPD) state to radiation pressure dominated (RPD) state. The calculated magnetic field range was found to be approximately (3-9)$\times$ 10$^{12}$ G, with the lower limit being more likely. Our magnetic field value is also consistent with the findings of \citet{2019ApJ...885...18J}, where the significant broadening of the iron line at higher luminosities indicated the possibility of a lower magnetospheric radius, and hence a relatively low magnetic field ranging from 10$^{11}$ to 10$^{12}$ G.  \citet{2022MNRAS.516.1601B} also calculated the magnetic field using the reflection model \texttt{relxilllp} during the super-Eddington phase of the source. They found that during the peak of the outburst, the inner accretion radius was approximately 2-3$\times$10$^{7}$ cm, corresponding to a magnetic field of 3$\times$10$^{12}$ G. However, some studies have reported a magnetic field of the pulsar as an order of magnitude higher than the estimated value in our work and previous studies. For instance, \citet{2022ApJ...933L...3K} detected a cyclotron resonance scattering feature (CRSF) in the 120-146 keV range through phase-resolved spectroscopy during the bright phase of the outburst. The calculated magnetic field corresponding to this observation is approximately 1.6$\times$10$^{13}$ G, which is the strongest ever detected for a neutron star in binaries. This component is thought to represent the quadrupole component of the magnetic field.  

A noticeable break is also observed in the power density spectra, as shown in Figure~\ref{fig:QPO_detection}. The evolution of the obtained break frequency (Br$_{f}$) with luminosity is shown in panel (e) of Figure~\ref{fig:QPO_evolve_L}.  Below the luminosity of $\sim$7.5$\times$10$^{37}$ erg s$^{-1}$, the $Br_{f}$ remain around 80 mHz. However, the break frequency reached 140 mHz, exceeding the pulsar spin frequency (101 mHz) beyond a luminosity of 7.5$\times$10$^{37}$ erg s$^{-1}$. This kind of evolution was also observed by  \citealt{2020MNRAS.491.1857D} (Figure 6) in the 20-40 keV Insight-HXMT light curve. The important point to note is that the first spectral transition was also observed around 7.5$\times$10$^{37}$ erg s$^{-1}$, although the connection is not clear yet. Again, the slope of the power laws below and above the break frequency varies with luminosity between 0.5-1 and  2-3, respectively. Above mentioned slopes $\Gamma_{1}$ \& $\Gamma_{2}$ in the PDS are mentioned to be created by variabilities in the accretion disc and magnetosphere of the neutron star, respectively \citep{1993ApJ...411L..79H, 2009A&A...507.1211R}.  This also suggests the conversion of matter flow from accretion disc flow to magnetospheric flow.  The higher values of $\Gamma_{2}$ than $\Gamma_{1}$ indicates suppression of  variabilities within the magnetosphere \citep{2009A&A...507.1211R}. Also, the observed break frequency, being almost 1.4$\times$ the spin frequency, suggests that the variability timescale relative to the break frequency is not directly associated with the expected Keplerian timescale at the inner edge of the disc \citep{2009A&A...507.1211R}. This conclusion can be drawn particularly at low luminosity levels (as in our study), where the spin period is expected to be around the same duration.

\subsection{Spectral characteristics of Swift~J0243.6+6124}

To gain a detailed understanding of the X-ray emission mechanisms from the source and to complement our timing studies, we performed spectral analysis using high-cadence NICER data.  Our study covers a wide time range from 2017 to 2023 that included the giant outburst, multiple subsequent normal outbursts, and the recent 2023 X-ray outburst. During our spectral analysis, the highest luminosity of the source in the 0.7--10 keV range was estimated to be 1.53$\times$10$^{39}$ erg s$^{-1}$. This result signifies that during the observation period, the source exceeds the Eddington luminosity limit ($L_{Edd}$ = 1.25$\times$10$^{38}$ erg s$^{-1}$ for a typical 1.4M$\odot$ neutron star), the maximum possible luminosity for a spherically symmetric emitting source. We are observing luminosity beyond the Eddington limit because of its calculation assumptions: First, the accretion is spherical, and second, the incoming matter and outgoing photons interact through Thomson scattering. But in the case of highly magnetized neutron stars, the interaction cross-section can be much below the Thomson scattering cross-section, and the radiation can escape in the perpendicular direction to the mass accretion direction, which on aggregate can increase the luminosity of the pulsar beyond the Eddington luminosity \citep{1976MNRAS.175..395B}.

Our spectral analysis revealed two distinct transitional luminosities, denoted as $L_{1}$ and $L_{2}$, which are $\approx$7.5 $\times$ 10$^{37}$ and $\approx$2.1$\times$10$^{38}$ erg s$^{-1}$ in 0.7-10.0 keV energy range, respectively for a distance of 7 kpc. These transitions are observed in continuum parameters such as photon index and cutoff energy, as shown in Figures~\ref{fig:spec_evolve_L} \& ~\ref{fig:spec_evolve_L2}. Insight-HXMT observed the transition around 1.5 $\times$10$^{38}$ erg s$^{-1}$ and  4.4 $\times$10$^{38}$ erg s$^{-1}$ in the 2-250 keV band at a distance of 6.8 kpc \citep{2020ApJ...902...18K}. These two transitions in spectral parameter evolution are observed for the first time in Swift J0243.6+6124. This type of behavior has possibly not been observed before for any other accretion-powered pulsars. In a study conducted by \citet{2013A&A...551A...1R}, the spectral analysis of nine Be/X-ray binary pulsars revealed the presence of two branches, indicating a single-luminosity transition. However, the photon index of these sources evolved opposite to Swift~J0243.6+6124, i.e., the photon index decreased with an increase in luminosity up to the first transition point (Figure~6 of \citealt{2013A&A...551A...1R}) except for Swift~J1626.6-5156, where the photon index remained constant. This shows that the X-ray emission mechanisms in  Swift~J0243.6+6124 are distinct from other accretion-powered pulsars. \citet{2013A&A...551A...1R} only found a single transition point in the evolution of spectral parameters with luminosity, representing the presence of two accretion modes on either side of the transition point or critical luminosity. However, in the case of Swift~J0243.6+6124, we found two transition points that suggest the presence of three distinct accretion modes, implying more complex behavior in this source.

In accretion-powered X-ray pulsars, the photon index evolves with the luminosity depending on the accretion regimes, where the accretion column acts as the primary source of the X-ray emission. The pulsars usually show a negative correlation between the photon index and luminosity below a single transition point or critical luminosity. A positive correlation is expected between these two parameters in the super-critical luminosity domain. In the sub-critical luminosity regime, the shock region in the column is much closer to the neutron star surface \citep{2012A&A...544A.123B}. The shock region moves up when the mass accretion rate increases, contributing to a taller column in the super-critical luminosity domain. In the latter case, the photons could not acquire enough energy through bulk Comptonization to produce high-energy photons, leading to a softer spectrum with increasing luminosity. On the other hand, in the sub-critical regime (below the transition point), the size of the interaction region decreases with luminosity. This leads to an increase in the optical depth, thereby hardening the spectrum. However, in the case of Swift~J0243.6+6124, we made a remarkable observation: there is a positive correlation between the photon index and the X-ray luminosity up to a transition point of approximately 7.5$\times$10$^{37}$  erg s$^{-1}$ (first transition zone). Beyond this point, the correlation becomes negative up to a certain threshold (2.1$\times$10$^{38}$ erg s$^{-1}$; second transition zone), and subsequently, we observed no correlation beyond the luminosity of 2.1$\times$10$^{38}$ erg s$^{-1}$.

{\bf First accretion mode ($L_{X}$ $\leq$ $L_{1}$):} We observe a positive correlation between the photon index and luminosity in the first transition zone below $\sim$7.5$\times$10$^{37}$  erg s$^{-1}$.  The presence of the observed softer spectrum indicates that lower energy X-ray photons are contributing more to the pulsar energy continuum in comparison to bulk Comptonization of photons with infalling electrons in the accretion column. The softer spectral shape may arise due to blackbody emission from the source. We observed a rise in the soft blackbody ($kT$ $\sim$ 0.1-0.7 keV) photon emitting region of size ($Radius$) $\sim$ 10-30 km in this accretion mode. The observed temperature and size of the emitting region suggest a potential combination of photons from both the neutron star's surface and the hot spot region \citep{2019MNRAS.488.4427Z, Elshamouty_2016}. This type of soft photon emitting region of temperature between 0.3-0.4 keV and size 25-38 km was also identified by \citealt{2021MNRAS.500..565B} using AstroSat during this luminosity range. Moreover, earlier studies have revealed the presence of a single \citep{2018MNRAS.474.4432J, 2020ApJ...902...18K} or multiple \citep{2019ApJ...873...19T} blackbody components representing the soft X-ray emission from the column and/or optically thick outflow in the spectrum of Swift~J0243.6+6124. The presence of these thermal regions may contribute to a softer spectrum. 
 Again as the X-ray luminosity increases, the gas shock region (photon emitting region) is expected to rise within the polar region (panel (b) of Figure 1 in \citealt{2012A&A...544A.123B}). At this stage, the temperature of the plasma increases in the shocked region. Consequently,  the shock height rises with mass accretion rate, resulting in a lower optical depth and less efficient cooling of the accreting plasma. As a consequence, the plasma temperature in the accretion column rises, explaining the correlation between $PI$  and $E_{cut}$  in the first transition zone.

{\bf Second accretion mode ($L_{1}$ $\leq$ $L_{X}$ $\leq$ $L_{2}$): } In this regime, both the photon index and the cutoff energy exhibit a correlated behavior with each other, but both parameters show an anti-correlation with the luminosity. This suggests the presence of a different emission mechanism compared to the previous accretion mode. In the above prescribed luminosity range, the accretion column shows different properties which are also reflected in the timing analysis \citep{2020ApJ...902...18K}. This type of behavior has been previously observed in the sub-critical regime of several sources such as 4U~0115+63, 1A~0535+262, 1A~1118-612, GRO~J1008-57 \citep{2013A&A...551A...1R}, EXO~2030+375 \citep{Epili2017MNRAS.472.3455E,2021JApA...42...33J}, 2S~1417-624 \citep{2018MNRAS.479.5612G}, and SMC~X-2 \citep{2023MNRAS.521.3951J}. In this luminosity range, it is expected that a radiation shock is present in the accretion column, though its strength is not sufficient enough to bring the matter to a complete rest at the stellar surface \citep{1976MNRAS.175..395B,2012A&A...544A.123B}. Instead, Coulomb interactions near the base of the accretion column play a significant role in reducing the plasma velocity \citep{1991ApJ...367..575B,1993ApJ...418..874N}. As a result, the height of the emitting region decreases with increasing luminosity following the relation of h$_{e}$ $\propto$ $L^{-5/7}_{X}$ (Equation~51 in \citealt{2012A&A...544A.123B}). The reduction in the size of the sinking region, or the Comptonization region, leads to an increase in optical depth, resulting in the production of harder photons and thus a lower photon index \citep{2012A&A...544A.123B}. Additionally, the cutoff energy decreases as the cooling mechanism through Comptonization dominates over the heating mechanism in this luminosity range, given the increased density resulting from the decreasing height of the emission region.

{\bf Third accretion mode ($L_{X}$ $\geq$ $L_{2}$):} In this luminosity range, the source is in a super-critical state, and we found that the $PI$ remains almost constant with luminosity, while the $E_{cut}$ value decreases. The super-critical state is characterized by the dominance of radiation pressure in controlling the overall flow dynamics of the plasma, ultimately causing the plasma to come to rest on the stellar surface \citep{1973NPhS..246....1D,1976MNRAS.175..395B}. The height of the emission region in this case increases with luminosity as h$_{e}$ $\propto$ $L_{X}$ (Equation~40, \citealt{2012A&A...544A.123B}). In this regime, the effective velocity of the incoming electrons decreases due to the balanced advection (inwards) and diffusion (outwards) of photons \citep{2012A&A...544A.123B}. As a consequence of the low effective electron velocity, photons do not acquire sufficient energy through bulk Comptonization, leading to spectral softening. Moreover, in this state, we detected a blackbody emission with a temperature of $\sim$0.1 keV and size of $\sim$200 km. The blackbody with such characteristics may indicate the presence of optically thick outflow during the super-Eddington phase as suggested by \citet{2019ApJ...873...19T} \& \citet{2021MNRAS.500..565B}.

In our spectral analysis within the energy range of 0.7-10 keV, in addition to the evolving iron emission lines (see e.g., \citealt{2019ApJ...885...18J}), we also observed the presence of emission lines around 1 keV (Figure \ref{fig:114id}).   These emission lines exhibited a detectable intensity during the super-Eddington phase. There is a low probability that the emission line is of instrument origin as suggested by the instrument team\footnote{\url{https://heasarc.gsfc.nasa.gov/docs/nicer/analysis_threads/arf-rmf/}}. The $\sim$1 keV line has been also detected in several sources with NICER such as  Serpens X-1~\citep{2018ApJ...858L...5L}, IGR J17062-6143 \citep{2021ApJ...912..120B}, and NGC 300 X-1~\citep{2022ApJ...940..138N}. The emission line displayed a distinctive single-peak shape, which led us to model it using a Gaussian function. The variations of the key parameters associated with these lines are presented in Figure~\ref{fig:0p7_1keV_line}.   Several other studies also reported the presence of this line in ULX pulsars ( \citealt{2023ApJ...955..124K}~\& references therein). These line features may be originated from a blend of Fe-L emissions \citep{2019A&A...627A..51G}. The evolution of the 1 keV line parameters with luminosity is presented in Figure \ref{fig:0p7_1keV_line}.  The variation of central line energy with luminosity remained within the error bars. However, $\sigma$ and $EW$  initially increase with luminosity up to the first transitional luminosity $L_{1}$ ( $\sim$ 7.5$\times$10$^{37}$ erg s$^{-1}$), and then beyond that decreases with luminosity. The line width reaches up to 0.1 keV. As a result, the velocity of the line-emitting material can be approximated to be around 10\% of the speed of light based on Doppler broadening. This may suggest the lines might have originated from the accretion disc or from an ultra-fast outflow that has been suggested during the super-Eddington phase of the pulsar  \citep{2019MNRAS.487.4355V,2019ApJ...885...18J}.

\section{Conclusion}
\label{sec:6}
In conclusion, our timing and spectral analysis of the first ultraluminous X-ray source in our Galaxy, Swift~J0243.6+6124, using NICER data, provided valuable insights into its behavior during different luminosity phases. We observed a luminosity-dependent break in power density spectra, signaling changing accretion dynamics, and identified quasi-periodic oscillations within a specific luminosity range. During the 2023 outburst, the neutron star exhibited a spin-up state and variations in its pulse profile. Spectral analysis revealed two luminosity-dependent transitions at a luminosity of $L_{1}$ $\approx$ 7.5$\times$10$^{37}$ erg s$^{-1}$ and  $L_{2}$ $\approx$ 2.1 $\times$ 10$^{38}$ erg s$^{-1}$ in continuum parameters, highlighting three distinct accretion modes during the giant outburst. we detected a soft blackbody component ($kT$ $\sim$ 0.08-0.7 keV), which underwent a discontinuous transition as the source evolved from a sub-Eddington to a super-Eddington state. Notably, during the super-Eddington state, we observed emission lines around 1 keV, indicating X-ray reflection from the accretion disc or outflow material.

\section*{acknowledgments}
We thank the anonymous reviewer for the constructive suggestions that helped us to improve the manuscript. The research work at the Physical Research Laboratory is funded by the Department of Space, Government of India.  This research has made use of NICER mission data and X-ray data analysis software provided by the High Energy Astrophysics Science Archive Research Center (HEASARC), which is a service of the Astrophysics Science Division at NASA/GSFC. We acknowledge the use of public data from the Swift data archive.

\vspace{5mm}
\facilities{  ADS, HEASARC, NICER, Swift (BAT)}

\software{ HEASOFT (v6.30), \texttt{XSPEC} (v-12.11.0, \citet{1996ASPC..101...17A})    }

\bibliography{Swiftj0apj}{}
\bibliographystyle{aasjournal}

\end{document}